\documentclass{sig-alternate}
\usepackage[english]{babel}
\usepackage{graphicx}
\DeclareGraphicsExtensions{.eps,.pdf,.png,.jpg}

\begin{document}
%
\conferenceinfo{PRINF 2015}{Dresden, Germany}

\title{Option contracts for a privacy-aware market}
%
%
%
%
%

\numberofauthors{2} 
%
\author{
%
%
\alignauthor
Maurizio Naldi\\
       \affaddr{Dpt. of Computer Science and Civil Engineering}\\
       \affaddr{Via del Politecnico 1}\\
       \affaddr{Rome, Italy}\\
       \email{naldi@disp.uniroma2.it}
\alignauthor
Giuseppe D'Acquisto\\
       \affaddr{Dpt. of Computer Science and Civil Engineering}\\
       \affaddr{Via del Politecnico 1}\\
       \affaddr{Rome, Italy}\\
       \email{dacquisto@ing.uniroma2.it}
}

\maketitle
\begin{abstract}
Suppliers (including companies and individual prosumers) may wish to protect their private information when selling items they have in stock. A market is envisaged where private information can be protected through the use of differential privacy and option contracts, while privacy-aware suppliers deliver their stock at a reduced price. In such a marketplace a broker acts as intermediary between privacy-aware suppliers and end customers, providing the extra items possibly needed to fully meet the customers' demand, while end customers book the items they need through an option contract. All stakeholders may benefit from such a marketplace. A formula is provided for the option price, and a budget equation is set for the mechanism to be profitable for the broker/producer. 
\end{abstract}




\section{Introduction}
When dealing with privacy in a marketplace, the accent is often on customers' privacy, who may not wish to divulge many private details or even the items they are purchasing. Platforms have been devised to enforce customers' privacy requirements over the complete lifecycle of private data  \cite{karjoth2003platform}. A market for customers' privacy is expected to emerge \cite{rust2002customer}. Instead, little attention has been paid to the wish of suppliers to protect their private information. Typically a company may wish to select the level of information it provides to its customers, but the widespread adoption of e-shops divulges a lot of details about company's operations, not just to prospective customers but to everyone accessing the e-commerce platform, including competitors. In addition, suppliers may now be not just companies but individuals (prosumers) who wish to sell product they happen to own \cite{Vargo2006}.

However, companies may wish to keep their data (e.g., their level of stock for a given product) secret. At the same time individuals acting as suppliers may wish not to be profiled and keeping secret the products they happen to own. The definition of a marketplace where suppliers can sell their products while retaining privacy is then a relevant issue. 

In this paper, we claim that such a marketplace may be set up with benefits for all the stakeholders (a broker/producer, privacy-aware suppliers, and end customers) through the use of differential privacy mechanisms and option contracts subscribed by end customers. We define such a marketplace by identifying all the relevant cash flows. We provide a formula for the price of the option contract and set the budget equation for the broker for the mechanism to be profitable. 

The paper is organized as follows. The market is described in Section \ref{market}, while the option price is derived in Section \ref{opzioni} and employed to set an overall budget equation for the broker in Section \ref{budget}.

\section{Market definition}
\label{market}
Let's consider a database of suppliers where information can be obtained about the availability of a set of items, but suppliers are somewhat screened. Suppliers could be vendors whose typical line of business does not include those products or wish to get rid of some remainders, or individuals (prosumers) who happen to have those products in their availability. For example, the database could contain the number of items available for sale at each supplier, so that the vertical sum across all suppliers included in the database would tell us the overall number of items available. Such a database, providing statistics about the entities included in it, is called a statistical database \cite{shoshani1982}. However, in a statistical database releasing statistical information may compromise the privacy of individual contributors. But suppliers may wish not to divulge those information; for example they do not want competitors (who could access the database) to know their level of stock, or, as individuals, they do not wish to be profiled about the items they own. If suppliers wish to be screened, a curator may sit between the users, posing the query, and the database. The responses to these queries may be modified by the curator in order to protect the privacy of the contributors \cite{dwork2008}, for example so as not to tell us exactly either which supplier can provide us with those items or how many items in the set are available. Instead of providing the exact number, the database provides us with an obfuscated number, which is more or less close to the exact figure. A mechanism to achieve differential privacy is the use of noisy sums: the response to a counting query is the sum of the true figure and some noise \cite{dwork2011}. The use of a statistical database plus the use of noisy sums may therefore protect the private information of suppliers.

When end customers demand for a number of items, the uncertainty surrounding the actual availability of those items doesn't allow  to close deals. In the presence of such privacy constraints, we postulate that a market can develop through the introduction of a broker/producer and the use of option contracts.
    
Let's consider the case where end customers demand for $k^{*}$ items. A broker commits to provide them with the number of items required. In fact, the broker may procure those items either by producing them itself (at a unit production cost $c_{p}$) or by resorting to \textit{privacy-aware suppliers}, whose availability is known through the statistical database previously mentioned. As already said, privacy-aware suppliers do not release full information about the availability of their products, but release instead an obfuscated number $\hat{k}$, which is generally different from the true number $k$ of items that they can provide (though the broker may obtain a refined estimate of the true number through Bayesian analysis \cite{naldi2014differential}).  

The privacy enjoyed by privacy-aware suppliers is reflected in the price $c_{s}$ they advertise. Prices set by privacy-aware suppliers depend on the level of obfuscation (i.e. privacy protection): the higher the level of obfuscation (embodied by the variance of the added noise), the lower the price. Assuming $c_{s}<c_{p}$, the broker has a real advantage to procure as many items as it can through privacy-aware suppliers at the reduced price $c_{s}$, and transfer part of that benefit to end customers by setting a lower end price. If the availability of items is not enough to satisfy the demand ($k<k^{*}$), the broker/producer produces the remaining items (but does not enjoy the full benefit of the reduced price).

\begin{figure}[htbp]
\begin{center}	
  \includegraphics[width=.85\columnwidth]{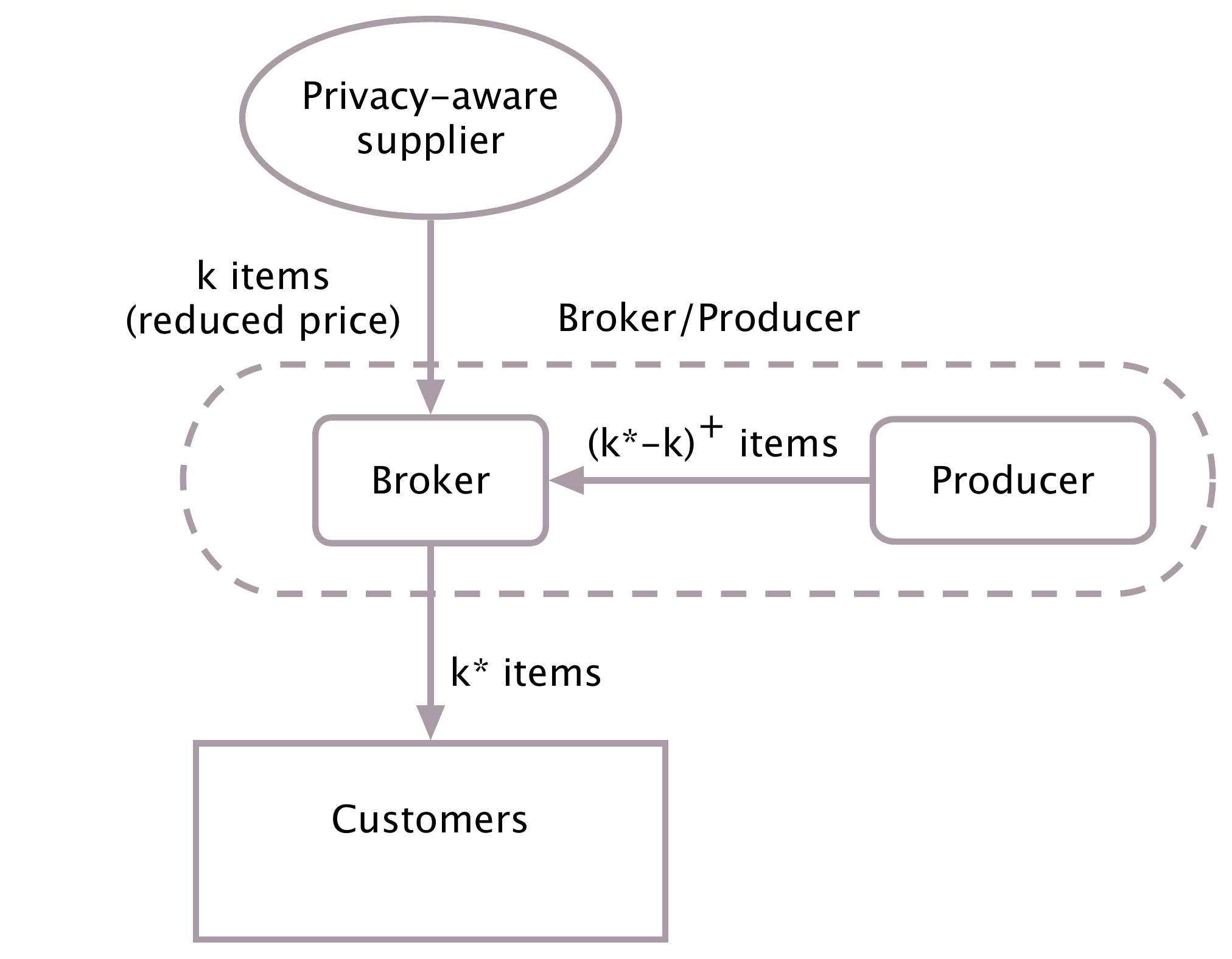}
	\caption{Supply relationships}
	\label{fig:sup}
\end{center}
\end{figure}

In order to exploit the offer by privacy-aware suppliers, the broker submits a query to the statistical database containing information about the availability of items and pays a fixed amount $c_{q}$ and receives the noisy response $\hat{k}$.  It commits to buy all the $k$ items available, though they may exceed the actual demand $k^{*}$. When the actual number of available items is disclosed (at delivery), it pays the privacy-aware suppliers the overall amount $c_{s}k$. If the demand is fully met ($k>k^{*}$) the broker does not have to produce any item; otherwise, the broker has to produce $k^{*}-k$ items at the unit cost $c_{p}$. The resulting supply chain is shown in \figurename~\ref{fig:sup}.

However, such a procedure is not free of risks for the broker/producer, which, on the one hand commits to provide its customers with the required items, but on the other hand is subject to the uncertainty determined by the unknown availability of items delivered by privacy-aware suppliers, with the risks deriving from the commitments to buy all the items available and, if required, to produce the remaining ones at a higher cost.

The broker/producer has therefore to hedge against such risks. A way we suggest is to resort to option contracts, which are described in the next section.

\section{Option contracts}
\label{opzioni}
As seen in the previous section, the broker/producer undergoes a risk when resorting to privacy-aware suppliers in order to meet customers' demand at a reduced cost, which it transfers to end customers through a reduced price. It needs however to hedge against such a risk. In this section we describe a mechanism, based on option contracts, by which it can achieve such protection.

Since the stakeholders that ultimately benefit from resorting to privacy-aware suppliers are end customers, the broker/producer may transfer some of that risk to them, asking them to pay a price to get the right to buy the desired number of items at a predetermined lower price  (i.e., a booking fee). In the language of financial markets, this is a \textit{call} option, since it endows the end customer with the right to buy \cite{hull2006options}. End customers are then required to subscribe a call option to be sure to get the right number of items the wish at a reduced price.

\begin{figure}[htbp]
\centering
  \includegraphics[width=.95\columnwidth]{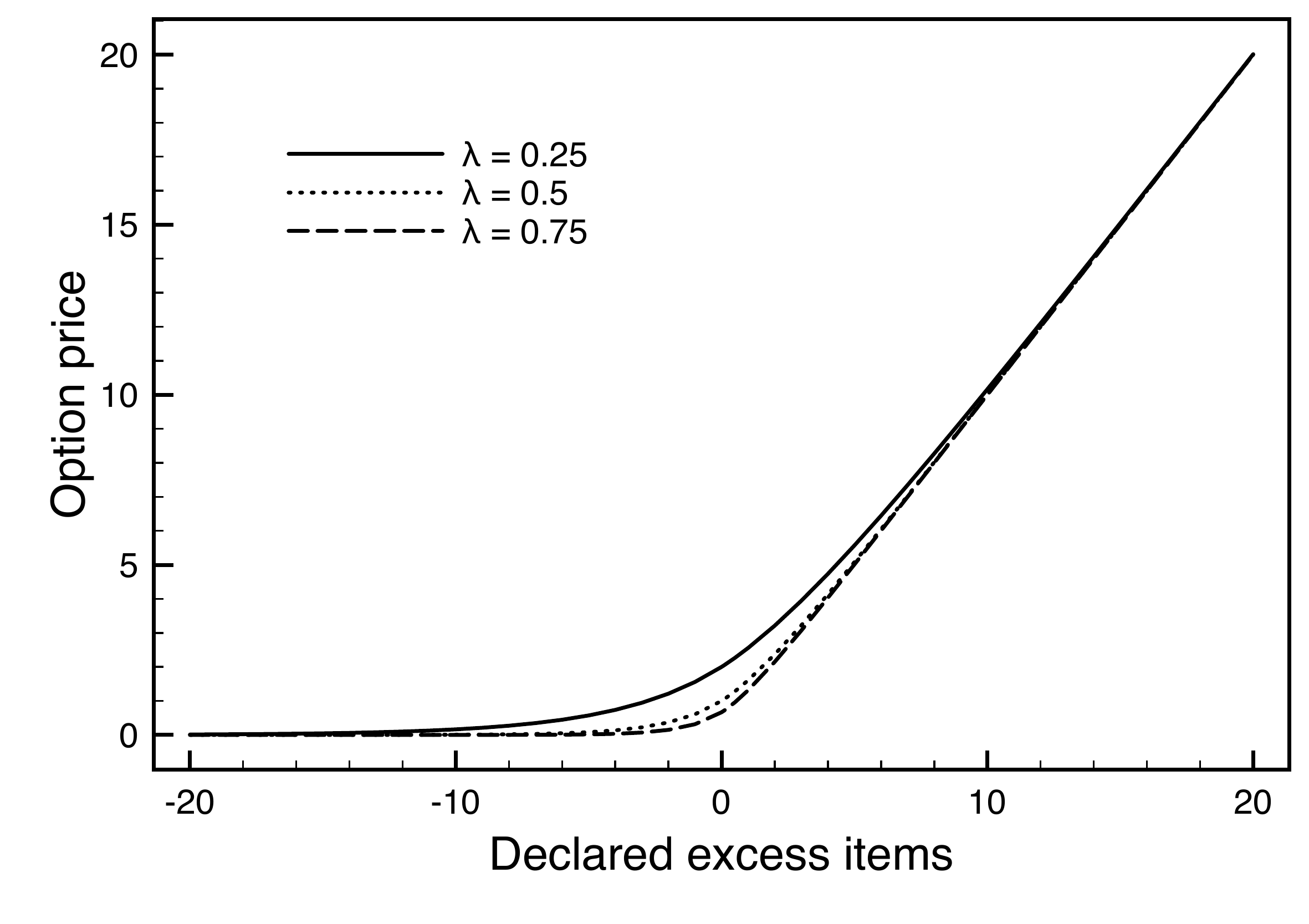}
	\caption{Option price}
	\label{fig:price}
\end{figure}

A critical issue in option contract is setting the right price. In the typical scenario, the amount to be paid for the option contract is expected to depend on the current value of the items for sale, the predetermined price to be paid if the option is exercised, and the expected behaviour of the item's value in the period from the option contract underwriting to the exercise time. A simple form of pricing is given by the Black-Scholes formula \cite{davis2010}, but a form tailored for the context is to be derived here. In \cite{MNcns15}, the risk of having to buy the items exceeding the demand has been analysed, and the following pricing formula has been derived for the case where Laplacian noise is added to form the noisy sum \cite{Sarathy2011}: 
\begin{equation}
\label{optprice}
p_{\textrm{opt}} = \mathbb{E}_{k}\left[(k-k^{*})^{+}c_{\textrm{s}}\vert \hat{k}\right] =  c_{\textrm{s}}\left[(\hat{k}-k^{*})^{+}+\frac{1}{2\lambda}e^{-\lambda \vert \hat{k} - k^{*}\vert}\right],
\end{equation}
where $\lambda$ is the shape parameter of the Laplace distribution: the smaller $\lambda$, the greater the differential privacy. This expression considers just the risk transfer concerning the extra-cost of buying the excess items provided by privacy-ware suppliers and does not consider the end price paid by customers.
As can be seen in Equation (\ref{optprice}), the price is basically the cost of the actual number of excess items plus a term that accounts for the introduction of noise in the database response and vanishes as $\lambda$  grows towards infinity (i.e., as the level of privacy reduces) and the declared number of available items gets farther from the demand. In \figurename~\ref{fig:price}, we can see how the option price moves for three different values of $\lambda$ and $c_{s}=1$. The difference with respect to the cost of the actual number of excess items is significant just when the number of declared items is close to the demand.

\begin{figure}[htbp]
\centering
  \includegraphics[width=.95\columnwidth]{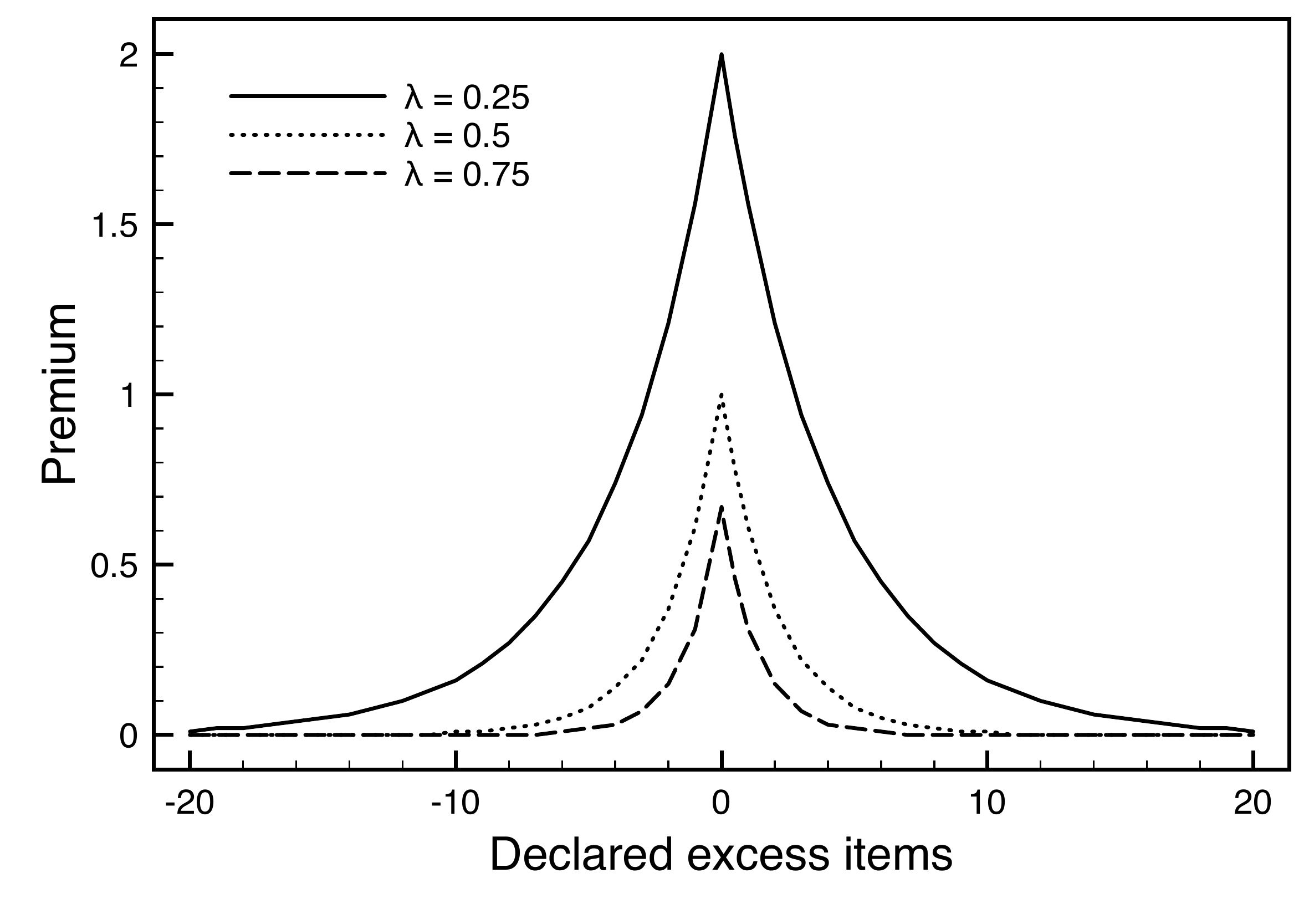}
	\caption{Premium}
	\label{fig:premium}
\end{figure}

We can consider that difference as the premium to be paid to cover the uncertainty introduced to Laplacian noise. After defining the premium
\begin{equation}
X = c_{s}\left[ \mathbb{E}_{k}\left(k-k^{*}\right)^{+} - \left(k-k^{*}\right)^{+}\right]
\end{equation}
we show it in \figurename~\ref{fig:premium} for the same case reported in \figurename~\ref{fig:price}. We see that the premium gets its maximum when the number $\hat{k}$ of declared items is exactly equal to the demand: the impact of uncertainty, and the protection from risk, is more expensive when the noisy response appears to exactly fit the demand.

\section{Broker's budget}
\label{budget}
The option price reported in Section \ref{opzioni} is just of the economic components in the interaction between the broker, the privacy-aware suppliers and the end customers. In order to examine the profitability of the market mechanism for the broker/producer, we must get an overall view. In this section, we review all the cash flows concerning the broker/producer and write a budget equation.

The broker's expenses are represented by:
\begin{itemize}
\item a fixed contribution $c_{\textrm{q}}$ for the query, depending on the noise variance;
\item a unit price $c_{\textrm{s}}$ for each item supplied by privacy-aware suppliers
\item an internal unit cost $c_{\textrm{p}}$ for each item it must produce to meet the demand if the items delivered by privacy-aware suppliers are not enough.
\end{itemize}

On the other hand, the revenues are:
\begin{itemize}
\item the option price $p_{\textrm{opt}}$, for letting its end customer the opportunity to benefit from the lower priced items offered by privacy-aware suppliers;
\item the unit price $p_{\textrm{s}}$ for each item delivered to end customers.
\end{itemize}

The overall profit $B$ for the producer is then
\begin{equation}
B = p_{\textrm{opt}} + k^{*}\cdot p_{\textrm{s}}-c_{\textrm{q}} - k\cdot c_{\textrm{s}} - (k^{*}-k)^{+}c_{\textrm{p}}
\end{equation}
 
Among the items entering the budget equation, we have already derived in Section \ref{opzioni} a preliminary expression for the option price (which does not take into account the end price $p_{s}$). As to the query cost $c_{q}$, for the time being we consider it as a fixed quantity, though for linear queries arbitrage-free pricing mechanisms have been proposed in \cite{Suciu2013}. Similarly, though in the following we assume $c_{s}$ to be fixed, we expect it to be a function of the level of differential privacy.
 
If the broker knew in advance the actual number of items delivered by privacy-aware suppliers, it could set the end price $p_{s}$ so as to strike a profit:
\begin{equation}
\label{minprice}
p_{s} > \frac{c_{\textrm{q}} + k\cdot c_{\textrm{s}} + (k^{*}-k)^{+}c_{\textrm{p}} - p_{\textrm{opt}} }{k^{*}} 
\end{equation}
Unfortunately that's not the case, and Equation (\ref{minprice}) contains a stochastic component due to $k$. 

If we look for a price capable of delivering a profit on the average, and recalling Equation (\ref{optprice}), the end price can be computed as follows (where, for simplicity, we have omitted the conditioning of all expected values on the declared number $\hat{k}$)
\begin{equation}
\label{minpriceav}
\begin{split}
p_{s} &> \frac{c_{\textrm{q}} + \mathbb{E}_{k}[k]\cdot c_{\textrm{s}} +\mathbb{E}_{k}[ (k^{*}-k)^{+}]c_{\textrm{p}} - p_{\textrm{opt}} }{k^{*}}\\
& =  \frac{c_{\textrm{q}} + \hat{k}\cdot c_{\textrm{s}} + \mathbb{E}_{k}[ (k^{*}-k)^{+}]c_{\textrm{p}}  - \mathbb{E}_{k}\left[(k-k^{*})^{+}\right]c_{\textrm{s}}  }{k^{*}}
\end{split}
\end{equation}

Again assuming a Laplace distribution for the added noise as in the computation of the option price, we can obtain
\begin{equation}
\mathbb{E}_{k}[ (k^{*}-k)^{+}] = \left( k^{*}-\hat{k} \right)^{+} + \frac{e^{-\lambda\vert \hat{k}-k^{*}  \vert}}{2\lambda}-\frac{e^{-\lambda\hat{k}}}{2}\left( k^{*}+\frac{1}{\lambda} \right)
\end{equation}
Introducing this expression in Equation (\ref{minpriceav}) and rearranging terms, we obtain the final expression
\begin{equation}
\label{minpriceav2}
\begin{split}
p_{s} > &\frac{c_{\textrm{q}} + \hat{k}\cdot c_{\textrm{s}} + c_{p}\left( k^{*}-\hat{k} \right)^{+} - c_{\textrm{s}}\left(\hat{k}-k^{*}\right)^{+} }{k^{*}}\\
& + \frac{\left( c_{p}-c_{s}\right)\frac{e^{-\lambda\vert \hat{k}-k^{*}  \vert}}{2\lambda} - \frac{c_{p}}{2}e^{-\lambda\hat{k}} \left( k^{*}+\frac{1}{\lambda} \right)  }{k^{*}}
\end{split}
\end{equation}

In order to obtain a practical mechanism to set the end price, we expect to introduce the dependence of the query price $c_{q}$ and the privacy-aware suppliers' price $c_{s}$ on the level of differential privacy.

\section{Benefits}
All the stakeholders are expected to benefit from the existence of such a market.

Privacy-aware suppliers are able to sell their products protecting their privacy at the same time. Though they are expected to sell at a lower price, they can decide the obfuscation level (i.e. the level of protection of their privacy) and establish the desired trade-off between privacy and profits.

On the other hand, the broker/producer can benefit from its dual role. As a producer it may sell its products at the current price but obtain an additional stream of revenue by selling option rights. As a broker, it may exploit the option mechanism and provide its end customers with the required number of items while taking advantage of the reduced price offered by privacy-aware suppliers. In a profitable context, the broker is incentivized to be trustable and keep the secret on the actual number of available items when it is disclosed at delivery.

Finally, end customers are certain to obtain the number of items they want, while taking advantage of privacy-aware price reductions at the same time. 

\section{Conclusions}
A marketplace has been envisaged where suppliers may retain their private information (e.g., their identity, and the type and quantity of items they stock). A broker acts between end customers and privacy-aware suppliers, employing option contracts to guarantee the full delivery of items. A formula has been provided for the option price, protecting the broker against the risk of buying items in excess of those demand by end customers. The option price is the cost of the actual number of excess items plus a term that accounts for the introduction of noise in the database response and vanishes as the level of privacy reduces  and the declared number of available items gets farther from the true one. The budget equation for the broker has also been set, and a formula has been derived for the minimum price to get the mechanism profitable for the broker on the average.
\balancecolumns



%
\bibliographystyle{abbrv}
\bibliography{Bib-privacy}  

\begin{thebibliography}{10}

\bibitem{davis2010}
M.~H. Davis.
\newblock {Black--Scholes Formula}.
\newblock {\em Encyclopedia of Quantitative Finance}, 2010.

\bibitem{dwork2008}
C.~Dwork.
\newblock Differential privacy: A survey of results.
\newblock In {\em Theory and Applications of Models of Computation}, pages
  1--19. Springer, 2008.

\bibitem{dwork2011}
C.~Dwork.
\newblock A firm foundation for private data analysis.
\newblock {\em Communications of the ACM}, 54(1):86--95, 2011.

\bibitem{hull2006options}
J.~C. Hull.
\newblock {\em Options, futures, and other derivatives}.
\newblock Pearson Education, 2006.

\bibitem{karjoth2003platform}
G.~Karjoth, M.~Schunter, and M.~Waidner.
\newblock Platform for enterprise privacy practices: Privacy-enabled management
  of customer data.
\newblock In {\em Privacy Enhancing Technologies}, pages 69--84. Springer,
  2003.

\bibitem{Suciu2013}
C.~Li, D.~Y. Li, G.~Miklau, and D.~Suciu.
\newblock A theory of pricing private data.
\newblock In {\em Proceedings of the 16th International Conference on Database
  Theory}, ICDT '13, pages 33--44, New York, NY, USA, 2013. ACM.

\bibitem{naldi2014differential}
M.~Naldi and G.~D'Acquisto.
\newblock Differential privacy for counting queries: can {B}ayes estimation
  help uncover the true value?
\newblock {\em arXiv preprint arXiv:1407.0116}, 2014.

\bibitem{MNcns15}
M.~Naldi and G.~D'Acquisto.
\newblock Option pricing in a privacy-aware market.
\newblock In {\em IEEE Conference on Communications and Network Security
  (CNS)}, Florence, Sept. 28-30, 2015.

\bibitem{rust2002customer}
R.~T. Rust, P.~Kannan, and N.~Peng.
\newblock The customer economics of internet privacy.
\newblock {\em Journal of the Academy of Marketing Science}, 30(4):455--464,
  2002.

\bibitem{Sarathy2011}
R.~Sarathy and K.~Muralidhar.
\newblock Evaluating {L}aplace noise addition to satisfy differential privacy
  for numeric data.
\newblock {\em Transactions on Data Privacy}, 4(1):1--17, 2011.

\bibitem{shoshani1982}
A.~Shoshani.
\newblock Statistical databases: Characteristics, problems, and some solutions.
\newblock In {\em Proceedings of the 8th International Conference on Very Large
  Data Bases}, pages 208--222. Morgan Kaufmann Publishers Inc., 1982.

\bibitem{Vargo2006}
S.~Vargo and R.~Lusch.
\newblock {\em {The service Dominant Logic. Debate and Directions}}.
\newblock Sharpe, 2006.

\end{thebibliography}
%
%
\end{document}